\title{NARROW ESCAPE, PART I}
\author{A. Singer \thanks{Department of Applied Mathematics,
Tel-Aviv University, Ramat-Aviv, 69978 Tel-Aviv, Israel, e-mail:
amits@post.tau.ac.il}\,,\ \ Z. Schuss\thanks{Department of
Mathematics, Tel-Aviv University, Tel-Aviv 69978,  Israel, e-mail:
schuss@post.tau.ac.il.}\,,\ \ D. Holcman\thanks{Department of
Mathematics, Weizmann Institute of Science, Rehovot 76100 Israel,
e-mail holcman@wisdom.weizmann.ac.il. }
\thanks{Keck Center, department of Physiology, UCSF, 513 Parnassus
Ave, San Francisco 94143 USA, e-mail holcman@phy.ucsf.edu.}\,,\ \
R. S. Eisenberg\thanks{Department of Molecular Biophysics and
Physiology, Rush Medical Center, 1750 Harrison St., Chicago, IL
60612, email: beisenbe@rush.edu }}
\documentclass[12pt]{article}
\usepackage{float}
\usepackage{amsmath}
\usepackage{epsfig, graphicx} 
\newcommand{\mb}[1]{\mbox{\boldmath$#1$}}
\newcommand{\p}{\partial}
\newcommand{\ds}{\displaystyle}
\newcommand{\beq}{\begin{eqnarray}}
\newcommand{\beqq}{\begin{eqnarray*}}
\newcommand{\eeq}{\end{eqnarray}}
\newcommand{\eeqq}{\end{eqnarray*}}
\newcommand{\x}{\mbox{\boldmath$x$}}
\newcommand{\y}{\mbox{\boldmath$y$}}

\font\bb=msbm10 at 12pt

\def\rR{\hbox{\bb R}}

\def\ds#1{\displaystyle{#1}}

\begin{document}
\numberwithin{equation}{section}
\maketitle
\begin{abstract}
\vspace{3mm}
\noindent A Brownian particle with diffusion coefficient $D$ is
confined to a bounded domain of volume $V$ in $\rR^3$ by a
reflecting boundary, except for a small absorbing window. The mean
time to absorption diverges as the window shrinks, thus rendering
the calculation of the mean escape time a singular perturbation
problem. We construct an asymptotic approximation for the case of
an elliptical window of large semi axis $a\ll V^{1/3}$ and show
that the mean escape time is $E\tau\sim\ds{\frac{V}{2\pi Da}}
K(e)$, where $e$ is the eccentricity of the ellipse; and
$K(\cdot)$ is the complete elliptic integral of the first kind. In
the special case of a circular hole the result reduces to Lord
Rayleigh's formula $E\tau\sim\ds{\frac{V}{4aD}}$, which was
derived by heuristic considerations. For the special case of a
spherical domain, we obtain the asymptotic expansion
$E\tau=\ds{\frac{V}{4aD}} \left[1+\frac{a}{R} \log \frac{R}{a} +
O\left(\frac{a}{R}\right) \right]$. This problem is important in
understanding the flow of ions in and out of narrow valves that
control a wide range of biological and technological function.
\end{abstract}
\section{Introduction}
We consider the exit problem of a Brownian motion from a bounded
domain, whose boundary is reflecting, except for a small absorbing
window. The mean first passage time to the absorbing window
(MFPT), $E\tau$, is the solution of a mixed Neumann-Dirichlet
boundary value problem (BVP) for the Poisson equation, known as
the corner problem, which has singularity at the boundary of the
hole \cite{Dauge}-\cite{Mazya2}. The MFPT grows to infinity as the
window size shrinks to zero, thus rendering its calculation a
singular perturbation problem for the mixed BVP, which we call the
narrow escape problem.
The narrow escape problem has been considered in the literature in
only a few special cases, beginning with Lord Rayleigh (in the
context of acoustics), who found the flux through a small hole by
using a result of Helmholtz \cite{Helmholtz}. He stated
\cite{Rayleigh} (p.176) ``{\em Among different kinds of channels
an important place must be assigned to those consisting of simple
apertures in unlimited plane walls of infinitesimal thickness. In
practical applications it is sufficient that a wall be very thin
in proportion to the dimensions of the aperture, and approximately
plane within a distance from aperture large in proportion to the
same quantity.''} More recently, Rayleigh's result was shown to
fit the MFPT obtained from Brownian dynamics simulations
\cite{Berez}. Another result was presented in \cite{Holcman},
where a two-dimensional narrow escape problem was considered and
whose method is generalized here. A related problem is that of
escape from a domain, whose boundary is reflecting, except for an
absorbing sphere, disjoint from the reflecting part of the
boundary \cite[and references therein]{Pinsky}. It differs from
the narrow escape problem in that there is no singularity at the
boundary and there is no boundary layer.
The mixed boundary value problems of classical electrostatics
(e.g., the electrified disk problem \cite{Jackson}), elasticity
(punch problems), diffusion and conductance theory, hydrodynamics,
and acoustics were solved, by and large, for special geometries by
separation of variables. In axially symmetric geometries this
method leads to a dual series or to integral equations that can be
solved by special techniques \cite{Sneddon}-\cite{Vinogradov}. The
special case of asymptotic representation of the solution of the
corner problem for small Dirichlet and large Neumann boundaries
was not done for general domains. The first attempt in this
direction seems to be \cite{Holcman}.
The narrow escape problem does not seem to fall within the theory
of large deviations \cite{DZ}. It is different from Kolmogorov's
exit problem \cite{MS77} of a diffusion process with small noise
from an attractor of the drift (e.g., a stable equilibrium or
limit cycle) in that the narrow escape problem has no large and
small coefficients in the equation. The singularity of
Kolmogorov's problem is the degeneration of a second order
elliptic operator into a first order operator in the limit of
small noise, whereas the singularity of the narrow escape problem
is the degeneration of the mixed BVP to a Neumann BVP on the
entire boundary. There exist precise asymptotic expansions of
$E\tau$ for Kolmogorov's exit problem, including error estimates
(see, e.g., \cite{HTB}, \cite{Freidlin}), which show that the MFPT
grows exponentially with decreasing noise. In contrast, the narrow
escape time grows algebraically rather than exponentially, as the
window shrinks.
Our first main result is a derivation of the leading order term in
the expansion of the MFPT of a Brownian particle with diffusion
coefficient $D$, from a general domain of volume $V$ to an
elliptical hole of large semi axis $a$ that is much smaller than
$V^{1/3}$,
 \beq
 E\tau\sim\ds{\frac{V}{2\pi Da}} K(e),\label{E}
 \eeq
where $e$ is the eccentricity of the ellipse, and $K(\cdot)$ is
the complete elliptic integral of the first kind. In the special
case of a circular hole (\ref{E}) reduces to
 \beq
 E\tau\sim\ds{\frac{V}{4aD}}.\label{MR}
 \eeq
Equation (\ref{E}) shows that the MFPT depends on the shape of the
hole, and not just on its area. This result was known to Lord
Rayleigh \cite{Rayleigh}, who considered the problem of the
electrified disk (which he knew was equivalent to finding the flow
of an incompressible fluid through a channel and to the problem of
finding the conductance of the channel), who reduced the problem
to that of solving an integral equation for the flux density
through the hole. The solution of the integral equation, which
goes back to Helmholtz \cite{Helmholtz} and is discussed in
\cite{Lure}, is proportional to $(a^2-\rho^2)^{-1/2}$ in the
circular case, where $\rho$ is the distance from the center of the
hole \cite{Jackson}-\cite{Fabrikant1}.
Note that equations (\ref{E}) and (\ref{MR}) are leading order
approximations and do not contain an error estimate. We prove
(\ref{E}) by using the singularity properties of Neumann's
function for three-dimensional domains, in a manner similar to
that used in \cite{Holcman} for two-dimensional problems. The
leading order term is the solution of Helmholtz's integral
equation \cite{Helmholtz}.
Our second main result is a derivation of the second term and
error estimate for a ball of radius $R$ with a small circular hole
of radius $a$ in the boundary,
\begin{eqnarray}
E\tau=\ds{\frac{V}{4aD}} \left[1+\frac{a}{R} \log \frac{R}{a} +
O\left(\frac{a}{R}\right) \right].\label{B}
\end{eqnarray}
Equation (\ref{B}) contains both the second term in the asymptotic
expansion of the MFPT and an error estimate. We use Collins'
method \cite{Collins1,Collins2} of solving dual series of
equations and expand the resulting solutions for small
$\varepsilon=a/R$. The estimate of the error term, which turns out
to be $O(\varepsilon \log \varepsilon)$, seems to be a new result.
An error estimate for eq.(\ref{E}) for a general domain is still
an open problem. We conjecture that it is $O(\varepsilon \log
\varepsilon)$, as is the case for the ball.
If the absorbing window touches a singular point of the boundary,
such as a corner or cusp, the singularity of the Neumann function
changes and so do the asymptotic results. In three dimensions the
class of isolated singularities of the boundary is much richer
than in the plane, so the methods of \cite{NarrowEscapeIII} cannot
be generalized in a straightforward manner to three dimensions. We
postpone the investigation of the MFPT to windows at isolated
singular points in three dimensions to a future paper.
In Section \ref{sec:general} we derive a leading order
approximation to the MFPT for a general domain with a general
small window. The leading order term is expressed in terms of a
solution to Helmholtz's integral equation, which is solved
explicitly for an elliptical window. In Section
\ref{sec:formulation} we obtain two terms in the asymptotic
expansion of the MFPT from a ball with a circular window and an
error estimate. Finally, we present a summary and list some
applications in Section \ref{Summary}. This is the first paper in
a series of three, the second of which considers the narrow escape
problem from a bounded simply connected planar domain, and the
third of which considers the narrow escape problem from a bounded
domain with boundary with corners and cusps on a two-dimensional
Riemannian manifold.
\section{General 3D bounded domain}
\label{sec:general}
A Brownian particle diffuses freely in a bounded domain
$\Omega\subset\rR^3$, whose boundary $\p\Omega$ is sufficiently
smooth. The trajectory of the Brownian particle, denoted $\x(t)$,
is reflected at the boundary, except for a small hole
$\p\Omega_a$, where it is absorbed. The reflecting part of the
boundary is $\partial \Omega_r =\partial \Omega -
\partial \Omega_a$. The lifetime of the particle in $\Omega$ is the first passage time
$\tau$ of the Brownian particle from any point $\x\in\Omega$ to
the absorbing boundary $\p\Omega_a$. The MFPT,
 \[v(\x)=E[\tau\,|\,\x(0)=\x],\]
is finite under quite general conditions \cite{Schuss}. As the
size (e.g., the diameter) of the absorbing hole decreases to zero,
but that of the domain remains finite, we assume that the MFPT
increases indefinitely. A measure of smallness can be chosen as
the ratio between the surface area of the absorbing boundary and
that of the entire boundary,
 $$\varepsilon =\ds{\frac{|\partial
 \Omega_a|}{|\partial \Omega|}} \ll 1,$$
(see, however, a pathological example in Appendix
\ref{Pathological}). The MFPT $v(\x)$ satisfies the mixed boundary
value problem \cite{Schuss}
\begin{eqnarray}
\Delta v(\mb{x}) &=& -\frac{1}{D}, \quad\mbox{for}\quad\x\in
\Omega,
\label{eq:v-general}\\
&&\nonumber\\
 v(\mb{x}) &=& 0, \; \quad\mbox{for}\quad\x\in
\partial\Omega_a,\label{eq:boundary-condition-v} \\
&&\nonumber\\
 \frac{\partial v(\mb{x})}{\partial n(\mb{x})} &=& 0,
\quad\mbox{for}\quad\x\in
\partial \Omega_r, \nonumber
\end{eqnarray}
where $D$ is the diffusion coefficient. According to our
assumptions $v(\x)\to\infty$ as the size of the hole decreases to
zero, e.g., as $\varepsilon\to0$, except in a boundary layer near
$\p\Omega_a$. Our purpose is to find an asymptotic approximation
to $v(\x)$ in this limit.
\subsection{The Neumann function and integral equations}
\label{sec:Neumann} To calculate the MFPT $v(\x )$, we use the
Neumann function $N(\x ,\mb{\xi})$ (see \cite{Holcman},
\cite{Pinsky}), which is a solution of the boundary value problem
\begin{eqnarray}
\label{eq:Neumann} \Delta_{\x } N(\x ,\mb{\xi}) & = &
-\delta(\x -\mb{\xi}),\quad\mbox{for}\quad \x ,\mb{\xi} \in \Omega, \\
&&\nonumber\\
 \frac{\partial N(\x ,\mb{\xi})}{\partial
n(\x )} & = & -\frac{1}{|\partial \Omega|},\quad\mbox{for}\quad \x
\in
\partial \Omega, \mb{\xi} \in \Omega, \nonumber
\end{eqnarray}
and is defined up to an additive constant. The Neumann function
has the form \cite{Garabedian}
\begin{equation}
\label{eq:Neumann-v} N(\x ,\mb{\xi}) = \frac{1}{4\pi|\x
-\mb{\xi}|} + v_S(\x ,\mb{\xi}),
\end{equation}
where $v_S(\x,\mb{\xi})$ is a regular harmonic function of
$\x\in\Omega$ and of $\mb{\xi}\in\Omega$. Green's identity gives
\begin{eqnarray}
& & \int_{\Omega} \left[N(\x ,\mb{\xi}) \Delta v(\x )  -
v(\x ) \Delta N(\x ,\mb{\xi}) \right] \,d\x   = \nonumber \\
&&\nonumber\\
& & = \int_{\partial \Omega} \left[ N(\x (\mb{S}),\mb{\xi})
\frac{\partial v(\x (\mb{S}))}{\partial n}-v(\x
(\mb{S}))\frac{\partial
N(\x (\mb{S}),\mb{\xi})}{\partial n} \right]\,dS  \nonumber \\
&&\nonumber\\
& & = \int_{\partial \Omega} N(\x (\mb{S}),\mb{\xi})
\frac{\partial v(\x (\mb{S}))}{\partial n}\,dS +\frac{1}{|\partial
\Omega|} \int_{
\partial \Omega} v(\x (\mb{S})) \,dS.  \nonumber
\end{eqnarray}
On the other hand, equations (\ref{eq:v-general}) and
(\ref{eq:Neumann}) imply that
\begin{equation}
\int_{\Omega} \left[N(\x ,\mb{\xi}) \Delta v(\x )  - v(\x ) \Delta
N(\x ,\mb{\xi}) \right] \,d\x   = v(\mb{\xi}) - \frac{1}{D}
\int_{\Omega} N(\x ,\mb{\xi})\,d\x , \nonumber
\end{equation}
hence
 \beq
&&v(\mb{\xi}) - \frac{1}{D} \int_{\Omega} N(\x ,\mb{\xi})
\,d\x  =\label{vx}\\
\nonumber\\
&& \int_{\partial \Omega} N(\x (\mb{S}),\mb{\xi}) \frac{\partial
v(\x (\mb{S}))}{\partial n}\,dS +\frac{1}{|\partial \Omega|}
\int_{
\partial \Omega} v(\x (\mb{S})) \,dS.\nonumber
 \eeq
Note that the second integral on the right hand side of
eq.(\ref{vx}) is an additive constant. Setting
 \beq
 C=\frac{1}{|\partial \Omega|}
\int_{
\partial \Omega} v(\x (\mb{S})) \,dS,\label{C}
 \eeq
we rewrite eq.(\ref{vx}) as
\begin{equation}
\label{eq:int-rep} v(\mb{\xi}) = \frac{1}{D} \int_{\Omega} N(\x
,\mb{\xi}) \,d\x  + \int_{\partial \Omega_a} N(\x
(\mb{S}),\mb{\xi}) \frac{\partial v(\x (\mb{S}))}{\partial n}\,dS
- C,
\end{equation}
which is an integral representation of $v(\mb{\xi})$. We define
the boundary flux density
\begin{equation}
g(\mb{S}) = \frac{\partial v(\x (\mb{S}))}{\partial n},
\end{equation}
choose $\mb{\xi} \in \partial \Omega_a$, and use the boundary
condition (\ref{eq:boundary-condition-v}) to obtain the equation
\begin{equation}
\label{eq:int-bound} 0 = \frac{1}{D} \int_{\Omega} N(\x ,\mb{\xi})
\,d\x  + \int_{\partial \Omega_a} N(\x (\mb{S}),\mb{\xi})
g(\mb{S}) \,dS - C,
\end{equation}
for all $\mb{\xi} \in \partial \Omega_a$. Equation
(\ref{eq:int-bound}) is an integral equation for $g(\mb{S})$ and
$C$.
To construct an asymptotic approximation to the solution, we note
that the first integral in equation (\ref{eq:int-bound}) is a
regular function of $\mb{\xi}$ on the boundary. Indeed, due to
symmetry of the Neumann function, we have from (\ref{eq:Neumann})
\begin{equation}
\label{eq:N-eq1} \Delta_{\mb{\xi}} \int_{\Omega} N(\x ,\mb{\xi})
\,d\x = -1\quad\mbox{for}\quad \mb{\xi} \in \Omega
\end{equation}
and
\begin{equation}
\label{eq:N-eq2} \frac{\partial}{\partial n(\mb{\xi})}
\int_{\Omega} N(\x ,\mb{\xi}) \,d\x  = -\frac{|\Omega|}{|\partial
\Omega |}\quad\mbox{for}\quad \mb{\xi} \in \p\Omega.
\end{equation}
Equation (\ref{eq:N-eq1}) and the boundary condition
(\ref{eq:N-eq2}) are independent of the hole $\p\Omega_a$, so they
define the integral as a regular function, up to an additive
constant, also independent of $\p\Omega_a$.
The assumption that for all $\x\in\Omega$, away from $\p\Omega_a$,
the MFPT $v(\x)$ increases to infinity as the size of the hole
decreases and eq.(\ref{C}) imply that $C\to\infty$ as as the size
of the hole decreases to zero. This means that for
$\mb{\xi}\in\p\Omega_a$ the second integral in
eq.(\ref{eq:int-bound}) must also become infinite in this limit,
because the first integral is independent of $\p\Omega_a$.
Therefore, the integral equation (\ref{eq:int-bound}) is to
leading order
\begin{equation}
\label{eq:g-int3} \int_{\partial \Omega_a} N(\x
(\mb{S}),\mb{\xi})) g_0(\mb{S}) \,dS =
C_0\quad\mbox{for}\quad\mb{\xi}\in\p\Omega_a,
\end{equation}
where $g_0(\mb{S})$ is the first asymptotic approximation to
$g(\mb{S})$ and $C_0$ is the first approximation to the constant
$C$. Furthermore, only the singular part of the Neumann function
contributes to the leading order, so we obtain the integral
equation
\begin{equation}
\label{eq:integral-equation} \frac{1}{2\pi} \int_{\p\Omega_a}
\frac{g_0(\x )}{|\x -\y |}\,dS_x = C_0,
\end{equation}
where $C_0$ is a constant,  which represents the first
approximation to the mean first passage time (MFPT). Note that the
singularity of the Neumann function at the boundary is twice as
large as it is inside the domain, due to the contribution of the
regular part (the ``image charge") and therefore the factor
$\ds{\frac{1}{4\pi}}$ of equation (\ref{eq:Neumann-v}) was
replaced by $\ds{\frac{1}{2\pi}}$. In general, the integral
equation (\ref{eq:integral-equation}) has no explicit solution,
and should be solved numerically.
\subsection{Elliptic hole}
\label{sec:elliptic}  When the hole $\partial\Omega_a$ is an
ellipse, the solution of the integral equation
(\ref{eq:integral-equation}) is known \cite{Rayleigh},
\cite{Lure}. Specifically, assuming the ellipse is given by
 \[\frac{x^2}{a^2}+\frac{y^2}{b^2}=1,\quad z=0,\quad(b\leq a),\]
the solution is
\begin{equation}
 \label{eq:flux-elliptic} g_0(\x ) =
\frac{\tilde g_0}{\sqrt{1-\ds{\frac{x^2}{a^2}-\frac{y^2}{b^2}}}},
\end{equation}
where $\tilde g_0$ is a constant (to be determined below). The
proof, originally given in \cite{Helmholtz}, is reproduced in
Appendix \ref{ap:lure}. To determine the value of the constant
$\tilde g_0$, we use the compatibility condition
\begin{equation}
\label{eq:comp} \int_{\p\Omega_a} g_0(\x ) \,dS_x =
\frac{|\Omega|}{D},
\end{equation}
obtained from the integration of eq.(\ref{eq:v-general}) over
$\Omega$. Using the value
\begin{equation}
\int_{\p\Omega_a} g_0(\x ) \,dS_x = \int_{-a}^a \, dx
\int_{-b\sqrt{1-\frac{x^2}{a^2}}}^{b\sqrt{1-\frac{x^2}{a^2}}}
\frac{\tilde
g_0\,dy}{\sqrt{1-\ds{\frac{x^2}{a^2}}-\ds{\frac{y^2}{b^2}}}} =
2\pi ab \tilde g_0
\end{equation}
and the compatibility condition (\ref{eq:comp}), we obtain
\begin{equation}
\tilde g_0 = \frac{|\Omega|}{2\pi Dab}.
\end{equation}
Hence, by equation (\ref{eq:K}), the leading order approximation
to $C$ is
\begin{equation}
C_0 = \frac{1}{2\pi}\int_{\p\Omega_a} \frac{g_0(\x )}{|\x -\y
|}\,dS_x =  \frac{|\Omega|}{2\pi D a} K(e),
\end{equation}
where $K(\cdot)$ is the complete elliptic integral of the first
kind, and $e$ is the eccentricity of the ellipse,
\begin{equation}
e = \sqrt{1-\frac{b^2}{a^2}}.
\end{equation}
In other words, the MFPT from a large cavity of volume $|\Omega|$
through a small elliptic hole is to leading order
\begin{equation}
E\tau(a,b)\sim \frac{|\Omega|}{2\pi Da} K(e).
\end{equation}
For example, in the case of a circular hole, we have  $e=0$ and
$K(0)=\ds{\frac{\pi}{2}}$,  so that
\begin{equation}
E\tau(a,a) \sim
\frac{|\Omega|}{4Da}=O\left(\frac1\varepsilon\right),\label{ETaa}
\end{equation}
provided
 \[\frac{|\Omega|^{2/3}}{|\p\Omega|}=O(1)\quad\mbox{for}\quad\varepsilon\ll1.\]
Equation (\ref{ETaa}) was used in \cite{Berez}, \cite{Weiss}. If
the mouth of the channel is not circular, the MFPT is different.
Equation (\ref{ETaa}) indicates that a Brownian particle that
tries to leave the domain ``sees'' finer details in the geometry
of the hole and the domain than just the quotient of the surface
areas.
The additional geometric features contained in the MFPT are
illustrated by the two interesting limits $e \ll 1$, where the
ellipse is almost circular, and $1-e \ll 1$, where the ellipse is
squeezed. In the case $e \ll 1$, we use the expansion of the
complete elliptic integral of the first kind \cite{Stegun}
\begin{equation}
K(e) = \frac{\pi}{2}\left\{1+\left(\frac{1}{2}\right)^2 e^2 +
\left(\frac{1\cdot 3}{2\cdot 4} \right)^2 e^4+\left(\frac{1\cdot 3
\cdot 5}{2\cdot 4 \cdot 6} \right)^3 e^6 + \cdots \right\}.
\end{equation}
In the second limit $1-e \ll 1$,  we find from the asymptotic
behavior \cite{Stegun}
\begin{equation}
\lim_{e\to 1} \left[K(e)-\frac{1}{2}\log \frac{16}{1-e} \right]=0
\end{equation}
that
\begin{equation}
E\tau \sim \frac{|\Omega|}{4\pi a}\log\frac{16}{1-e},
\quad\mbox{for}\quad 1-e\ll 1.
\end{equation}
The area of the hole is given by
\begin{equation}
S = \pi a b = \pi a^2 \sqrt{1-e^2},
\end{equation}
or equivalently
\begin{equation}
a = \frac{S^{1/2}}{\pi^{1/2}\left(1-e^2\right)^{1/4}},
\end{equation}
and the MFPT has the asymptotic form
\begin{equation}
E\tau \sim
\frac{\sqrt[4]{2}\,|\Omega|\left(1-e\right)^{1/4}}{4\sqrt{\pi
S}}\log\frac{16}{1-e},\quad\mbox{for}\quad 1-e\ll 1.
\end{equation}
\section{Explicit computations for the sphere}
\label{sec:formulation}
The analysis of Section \ref{sec:general} is not easily extended
to the computation, or even merely the estimation of the next term
in the asymptotic approximation of the MFPT. The explicit results
for the particular case of escape from a ball through a small
circular hole gives an idea of the order of magnitude of the
second term and the error in the asymptotic expansion of the MFPT.
If the domain $\Omega$ is a ball, the method of
\cite{Sneddon}-\cite{Fabrikant2}, \cite{Collins1}, and
\cite{Collins2} can be used to obtain a full asymptotic expansion
of the MFPT. We consider the motion of a Brownian particle inside
a ball of radius $R$. The particle is reflected at the sphere,
except for a small cap of radius $a=\varepsilon R$ and surface
area $4\pi R^2 \sin^2\ds{ \frac{\varepsilon}{2}}$, where it exits
the ball. We assume $\varepsilon\ll1$. The MFPT $v(r,\theta,\phi)$
satisfies the mixed boundary value problem for Poisson's equation
in the ball \cite{Schuss},
\begin{eqnarray}
 \Delta v(r,\theta,\phi) & = & -1,
\quad\mbox{for}\quad r<R, \quad 0\leq \theta \leq \pi,\quad 0 \leq
\phi < 2\pi,\nonumber \\
&&\nonumber\\
v(r,\theta,\phi)\bigg|_{r=R} & = & 0,\quad\mbox{for}\quad 0 \leq
\theta < \varepsilon, \quad 0
\leq \phi < 2\pi, \label{eq:Poisson} \\
&&\nonumber\\
 \frac{\partial v(r,\theta,\phi)}{\partial
r}\bigg|_{r=R} & = & 0, \quad\mbox{for}\quad \varepsilon \leq
\theta \leq \pi, \quad 0 \leq \phi < 2\pi,\nonumber
\end{eqnarray}
The diffusion coefficient has been chosen to be $D=1$.
Due to the cylindrical symmetry of the problem, the solution is
independent of the angle $\phi$, that is, $v(r,\theta,\phi) =
v(r,\theta)$, so the system (\ref{eq:Poisson}) can be written as
\begin{eqnarray*}
\Delta v(r,\theta) & = & -1, \quad \mbox{for}\quad r<R, \quad
0\leq \theta \leq \pi,
\nonumber \\
v(r,\theta)\bigg|_{r=R} & = & 0, \quad\mbox{for}\quad 0 \leq
\theta < \varepsilon, \nonumber\\
&&\nonumber\\
 \frac{\partial v(r,\theta)}{\partial r}\bigg|_{r=R}
& = & 0, \quad \mbox{for}\quad\varepsilon \leq \theta \leq \pi,
\end{eqnarray*}
where the Laplacian is given by
$$\Delta v(r,\theta) = \frac{1}{r^2}\frac{\partial}{\partial
r}\left(r^2 \frac{\partial v}{\partial r} \right) + \frac{1}{r^2
\sin \theta}\frac{\partial}{\partial \theta}\left(\sin \theta
\frac{\partial v}{\partial \theta} \right).$$
The function $\displaystyle f(r,\theta) = \frac{R^2-r^2}{6}$ is
the solution of the boundary value problem
\begin{eqnarray*}
\Delta f &=& -1, \quad \mbox{for}\quad r<R,\nonumber \\
&&\nonumber\\
 f \bigg|_{r=R} & = & 0.
\end{eqnarray*}
In the decomposition $v=u+f$, the function $u(r,\theta)$ satisfies
the mixed Dirichlet-Neumann boundary value problem for the Laplace
equation
\begin{eqnarray}
\Delta u(r,\theta) & = & 0, \quad \mbox{for}\quad r<R, \quad 0
\leq \theta \leq \pi,\nonumber\\
&&\nonumber\\
u(r,\theta)\bigg|_{r=R} & = & 0, \quad \mbox{for}\quad0 \leq
\theta < \varepsilon,\label{eq:bvp-u} \\
&&\nonumber \\
 \frac{\partial u(r,\theta)}{\partial r}\bigg|_{r=R}
& = & \frac{R}{3}, \quad\mbox{for}\quad \varepsilon \leq \theta
\leq \pi.\nonumber
\end{eqnarray}
Separation of variables suggests that
\begin{equation}
u(r,\theta) = \sum_{n=0}^\infty a_n \left(\frac{r}{R}\right)^n
P_n(\cos \theta),
\end{equation}
where $P_n(\cos\theta)$ are the Legendre polynomials, and the
coefficients $\{a_n\}$ are to be determined from the boundary
conditions
\begin{eqnarray}
\label{eq:boundary-1} u(r,\theta)\bigg|_{r=R} & = &
\sum_{n=0}^\infty
a_n P_n(\cos \theta) = 0, \quad 0 \leq \theta < \varepsilon, \\
&&\nonumber\\
 \label{eq:boundary-2} \frac{\partial
u(r,\theta)}{\partial r}\bigg|_{r=R} & = & \sum_{n=1}^\infty n a_n
P_n(\cos \theta) = \frac{R^2}{3}, \quad \varepsilon \leq \theta
\leq \pi.
\end{eqnarray}
Equations (\ref{eq:boundary-1}), (\ref{eq:boundary-2}) are dual
series equations of the mixed boundary value problem at hand, and
their solution results in the solution of the boundary value
problem (\ref{eq:bvp-u}). Dual series equations of the form
\begin{eqnarray}
\sum_{n=0}^\infty
a_n P_n(\cos \theta) & = & 0, \quad\mbox{for}\quad 0 \leq \theta < \varepsilon, \label{eq:sneddon-1}\\
&&\nonumber\\
 \sum_{n=0}^\infty (2n+1) a_n P_n(\cos
\theta) & = & G(\theta), \quad \mbox{for}\quad\varepsilon \leq
\theta \leq \pi \label{eq:sneddon-2}
\end{eqnarray}
are solved in \cite[eqs.(5.5.12)-(5.5.14), (5.6.12)]{Sneddon}.
However, the dual series equations
(\ref{eq:sneddon-1})-(\ref{eq:sneddon-2}) are different from
equations (\ref{eq:boundary-1})-(\ref{eq:boundary-2}). The factor
$2n+1$ that appears in equation (\ref{eq:sneddon-2}) is replaced
by $n$ in equation (\ref{eq:boundary-2}). What seems as a slight
difference turns out to make our task much harder. The factor
$2n+1$ fits much more easily into the infinite sums
(\ref{eq:sneddon-1})-(\ref{eq:sneddon-2}), because it is the
normalization constant of the Legendre polynomials.
\subsection{Collins' method} \label{sec:collins}
The solution of dual relations of the form (\ref{eq:boundary-2})
(see \cite[(5.6.19)-(5.6.20)]{Sneddon}) is discussed in
\cite{Collins1}, \cite{Collins2}. Specifically, assume that for
given functions $G(\theta)$ and $F(\theta)$ we have the
representation
\begin{eqnarray*}
\sum_{n=0}^\infty (1+H_n) b_n T_{m+n}^{-m}(\cos \theta) & = &
F(\theta),\quad\mbox{for}\quad \ 0 \leq \theta < \varepsilon, \\
&&\nonumber\\
 \sum_{n=0}^\infty (2n+2m+1) b_n T_{m+n}^{-m} (\cos
\theta) & = & G(\theta), \quad\mbox{for}\quad \varepsilon < \theta
\leq \pi,
\end{eqnarray*}
where $T_{m+n}^{-m}$ are Ferrer's associated Legendre polynomials
\cite{Erdelyi}, \cite{Roy} and $\{H_n\}$ is a given series that is
$O(n^{-1})$ as $n \rightarrow \infty$. Then for $m=0$, we have
\begin{eqnarray}
\label{eq:collins-1} \sum_{n=0}^\infty (1+H_n) b_n P_n(\cos
\theta) & = &
F(\theta), \quad\mbox{for}\quad 0 \leq \theta < \varepsilon, \\
&&\nonumber\\
 \label{eq:collins-2} \sum_{n=0}^\infty (2n+1) b_n
P_n (\cos \theta) & = & G(\theta), \quad\mbox{for}\quad
\varepsilon < \theta \leq \pi.
\end{eqnarray}
Setting $\displaystyle a_0=b_0, \ a_n = \frac{2n+1}{2n}b_n, \
n\geq1$ in equations (\ref{eq:boundary-1})-(\ref{eq:boundary-2})
results in
\begin{eqnarray}
\label{eq:b-1} \sum_{n=0}^\infty
(1+H_n) b_n P_n(\cos \theta) & = & 0, \quad \mbox{for}\quad0 \leq \theta < \varepsilon, \\
&&\nonumber\\
 \label{eq:b-2} \sum_{n=0}^\infty (2n+1) b_n P_n(\cos
\theta) & = & \frac{2R^2}{3}+b_0, \quad\mbox{for}\quad \varepsilon
\leq \theta \leq \pi.
\end{eqnarray}
Equations (\ref{eq:b-1})-(\ref{eq:b-2}) are equivalent to
(\ref{eq:collins-1})-(\ref{eq:collins-2}) with $\displaystyle
H_0=0, \ H_n = \frac{1}{2n},\ n\geq 1$, $F(\theta) = 0$, and
$\displaystyle G(\theta) = \frac{2R^2}{3}+b_0$. Collins' method of
solution consists in finding an integral equation for the function
 \beqq
h(\theta) = \sum_{n=0}^\infty (2n+1) b_n P_n(\cos \theta),
\quad\mbox{for}\quad 0 \leq \theta < \varepsilon,
 \eeqq
so that
 \beqq
\label{eq:b_n} b_n = \frac{1}{2} \int_0^{\varepsilon} h(\alpha)
P_n(\cos \alpha) \sin \alpha \, d\alpha + \frac{1}{2}
\int_{\varepsilon}^{\pi} G(\alpha) P_n(\cos \alpha) \sin \alpha \,
d\alpha.
 \eeqq
Substituting into equation (\ref{eq:collins-1}), with
$F(\theta)\equiv 0$, we find for $0 \leq \theta < \varepsilon$
that
\begin{eqnarray} \label{18}
0 & = & \frac{1}{2}\int_0^\varepsilon h(\alpha) \sum_{n=0}^\infty
(1+H_n) P_n(\cos
\alpha) P_n(\cos \theta) \sin \alpha \,  d\alpha \nonumber \\
&&\nonumber\\
 & & + \frac{1}{2}\int_\varepsilon^\pi G(\alpha)
\sum_{n=0}^\infty (1+H_n) P_n(\cos \alpha) P_n(\cos \theta) \sin
\alpha \, d\alpha.
\end{eqnarray}
\subsection{The asymptotic expansion} \label{subsec:strategy}
\noindent To facilitate the calculations, we consider first the
case $H_n=0$ for all $n$. Then we will show that the leading order
term obtained for this case is the same as that for the case
$H_n\neq 0$. In the latter case, we obtain the first correction to
the leading order term and an estimate on the remaining error.
\subsubsection{The leading order term when $H_n\equiv 0$} We will now
sum the series (\ref{18}) in the case $H_n\equiv 0$. First, we
recall Mehler's integral representation for the Legendre
polynomials \cite{Stegun}, \cite{Magnus},
\begin{equation}
P_n(\cos \theta) = \frac{\sqrt{2}}{\pi}\int_0^\theta
\frac{\cos(n+\frac{1}{2})u\,du}{\sqrt{\cos u - \cos \theta}},
\end{equation}
and the identity \cite{Sneddon}
\begin{equation}
\sqrt{2} \sum_{n=0}^\infty P_n(\cos \alpha)
\cos\left(n+\frac{1}{2}\right)u = \frac{H(\alpha-u)}{\sqrt{\cos u
- \cos \alpha}},
\end{equation}
where $H(x)$ is the Heaviside unit step function.  Then we obtain
for $u < \theta < \varepsilon < \alpha$,
\begin{eqnarray}
&& \frac{1}{2}\int_\varepsilon^\pi G(\alpha) \sum_{n=0}^\infty
P_n(\cos \alpha) P_n(\cos \theta) \sin \alpha \, d\alpha =
\nonumber \\
&&\nonumber\\
 & = &  \frac{1}{2}\int_\varepsilon^\pi G(\alpha)
\sum_{n=0}^\infty P_n(\cos \alpha)
\frac{\sqrt{2}}{\pi}\int_0^\theta
\frac{\cos(n+\frac{1}{2})u\,du}{\sqrt{\cos u - \cos \theta}} \sin
\alpha \, d\alpha \nonumber \\
&&\nonumber\\
 & = & \frac{1}{2\pi} \int_0^\theta
\frac{du}{\sqrt{\cos u - \cos \theta}} \int_\varepsilon^\pi
\frac{G(\alpha) \sin \alpha \, d\alpha}{\sqrt{\cos u - \cos
\alpha}}.\label{eq:52}
\end{eqnarray}
Similarly,
\begin{eqnarray}
&& \frac{1}{2}\int_0^\varepsilon h(\alpha) \sum_{n=0}^\infty
P_n(\cos \alpha) P_n(\cos \theta) \sin \alpha \, d\alpha =
\nonumber \\
&&\nonumber\\
 & = & \frac{1}{2\pi}\int_0^\theta
\frac{du}{\sqrt{\cos u - \cos \theta}} \int_u^\varepsilon
\frac{h(\alpha) \sin \alpha\,d\alpha}{\sqrt{\cos u - \cos
\alpha}}.\label{eq:53}
\end{eqnarray}
Hence,
 \beq
  &&\int_0^\theta \frac{du}{\sqrt{\cos u - \cos
\theta}} \int_u^\varepsilon \frac{h(\alpha) \sin
\alpha\,d\alpha}{\sqrt{\cos u - \cos \alpha}} =
\nonumber\\
&&\nonumber\\
&&-\int_0^\theta \frac{du}{\sqrt{\cos u - \cos \theta}}
\int_\varepsilon^\pi \frac{G(\alpha) \sin \alpha \,
d\alpha}{\sqrt{\cos u - \cos \alpha}}.\label{AbelT}
 \eeq
Equation (\ref{AbelT}) means that the Abel transforms
\cite{Whittaker} of two functions are the same, so that
\begin{equation}
\int_u^\varepsilon \frac{h(\alpha) \sin
\alpha\,d\alpha}{\sqrt{\cos u - \cos \alpha}} =
-\int_\varepsilon^\pi \frac{G(\alpha) \sin \alpha \,
d\alpha}{\sqrt{\cos u - \cos \alpha}},\label{318}
\end{equation}
because the Abel transform is uniquely invertible. Equastion
(\ref{318}) is an Abel-type integral equation, whose solution is
given by
\begin{equation}
h(\theta)\sin\theta = \frac{1}{\pi} \frac{d}{d\theta}
\int_{\theta}^\varepsilon \frac{\sin u \,du}{\sqrt{\cos \theta -
\cos u}}\int_\varepsilon^\pi \frac{G(\alpha) \sin \alpha \,
d\alpha}{\sqrt{\cos u - \cos \alpha}},
\end{equation}
or
\begin{equation}
\label{eq:h-H} h(\theta) = - \frac{2}{\sin\theta}\frac{d}{d\theta}
\int_{\theta}^\varepsilon \frac{H(u) \sin u \,du}{\sqrt{\cos
\theta - \cos u}},
\end{equation}
where
\begin{equation}
H(u) = -G(u,\varepsilon),
\end{equation}
and
\begin{equation}
G(u,\varepsilon) = \frac{1}{2\pi} \int_\varepsilon^\pi
\frac{G(\theta)\sin\theta\,d\theta}{\sqrt{\cos u - \cos \theta}}.
\end{equation}
The dual integral equations (\ref{eq:b-1})-(\ref{eq:b-2}) define
$G(\theta) = \displaystyle\frac{2R^2}{3}+b_0$, so that
\begin{eqnarray}
G(\psi,\phi)&=&\frac{1}{2\pi} \int_{\phi}^{\pi}
\left(\frac{2R^2}{3}+b_0\right) \frac{\sin\theta \, d
\theta}{\sqrt{\cos \psi - \cos \theta} }\label{eq:G}\\
&&\nonumber\\
&=& \left(\frac{2R^2}{3}+b_0\right)\frac{1}{\pi} \sqrt{\cos \psi -
\cos\theta}\,\bigg|_{\theta=\phi}^\pi \nonumber \\
&&\nonumber \\
 & =& \left(\frac{2R^2}{3}+b_0\right)\frac{1}{\pi}
\left(\sqrt{2}\cos\frac{\psi}{2}- \sqrt{\cos\psi - \cos
\phi}\right),\quad\mbox{for}\quad\psi < \phi.\nonumber
\end{eqnarray}
In particular, setting $n=0$ in equation (\ref{eq:b_n}) and using
equation (\ref{eq:h-H}), gives
\begin{eqnarray}
\label{eq:b_0} b_0 & =& \frac{1}{2} \int_0^{\varepsilon} h(\alpha)
\sin \alpha \, d\alpha + \frac{1}{2} \int_{\varepsilon}^{\pi}
\left(\frac{2R^2}{3}+b_0\right) \sin \alpha \, d\alpha\\
&&\nonumber\\
&=&\sqrt{2}\int_0^{\varepsilon} H(\psi)\cos \frac{\psi}{2} \,
d\psi + \left(\frac{2R^2}{3}+b_0\right)
\cos^2\frac{\varepsilon}{2}.\nonumber
\end{eqnarray}
Integrating equation (\ref{eq:G}), we obtain
\begin{eqnarray}
\label{eq:int-G} && \sqrt{2} \int_0^{\varepsilon}
G(\psi,\varepsilon)\cos
\displaystyle\frac{\psi}{2} \, d\psi =\\
&&\nonumber\\
&=&\left(\displaystyle\frac{2R^2}{3}+b_0\right)\displaystyle\frac{\sqrt{2}}{\pi}\int_0^{\varepsilon}
\left(\sqrt{2}\cos\displaystyle\frac{\psi}{2}- \sqrt{\cos\psi-\cos
\varepsilon}\right) \cos \displaystyle\frac{\psi}{2} \, d\psi \nonumber \\
&&\nonumber\\
& = &
\displaystyle\frac{\displaystyle\frac{2R^2}{3}+b_0}{\pi}(\varepsilon
+ \sin \varepsilon) -
\left(\displaystyle\frac{2R^2}{3}+b_0\right)\displaystyle\frac{4}{\pi}\int_0^{\displaystyle\sin
\displaystyle\frac{\varepsilon}{2}}
\displaystyle\frac{s^2\,ds}{\sqrt{\sin^2\displaystyle\frac{\varepsilon}{2}-s^2}}
\nonumber \\
&&\nonumber\\
& = &
\displaystyle\frac{\displaystyle\frac{2R^2}{3}+b_0}{\pi}(\varepsilon
+ \sin \varepsilon) -
\left(\displaystyle\frac{2R^2}{3}+b_0\right)\sin^2\displaystyle\frac{\varepsilon}{2}.\nonumber
\end{eqnarray}
Combining equations (\ref{eq:b_0}) and (\ref{eq:int-G}) gives
\begin{equation}
\label{eq:b_0-exact} b_0 =
\frac{2R^2}{3}\left(\frac{\pi}{\varepsilon + \sin \varepsilon}-1
\right)= \frac{2R^2}{3} \left(\frac{\pi}{2\varepsilon} +
O(1)\right)= \frac{|\Omega|}{4a}\left(1+
O\left(\frac{a}{R}\right)\right),
\end{equation}
where $\displaystyle |\Omega|=\frac{4\pi R^3}{3}$ is the volume of
the ball, and $a=R\varepsilon$ is the radius of the hole.
\subsubsection{The case $H_n \neq 0$}\label{Hn0}
The asymptotic expression (\ref{eq:b_0-exact}) for $b_0$, was
derived under the simplifying assumption that $H_n\equiv 0$.
However, we are interested in the value of $b_0$ which is produced
by the solution of the dual series equations
(\ref{eq:b-1})-(\ref{eq:b-2}), where $H_n=\ds{\frac{1}{2n}}$.
We sum the series (\ref{18}) by the identities
\begin{eqnarray}
&& \frac{1}{2}\int_0^\varepsilon h(\alpha) \sum_{n=0}^\infty H_n
P_n(\cos \alpha) P_n(\cos \theta) \sin \alpha \, d\alpha
\nonumber  \\
&&\nonumber\\ &= & \frac{1}{2}\int_0^\varepsilon h(\alpha)
\sum_{n=0}^\infty H_n \frac{\sqrt{2}}{\pi}\int_0^\alpha
\frac{\cos(n+\frac{1}{2})v\,dv}{\sqrt{\cos v - \cos \alpha}}
\frac{\sqrt{2}}{\pi}\int_0^\theta
\frac{\cos(n+\frac{1}{2})u\,du}{\sqrt{\cos u - \cos \theta}} \sin
\alpha \, d\alpha \nonumber \\
&&\nonumber\\
& = & \frac{1}{2\pi}\int_0^\varepsilon h(\alpha)\sin \alpha \,
d\alpha \int_0^\alpha  \frac{dv}{\sqrt{\cos v - \cos \alpha}}
\int_0^\theta \frac{K(u,v)\,du}{\sqrt{\cos u - \cos \theta}}  \nonumber \\
&&\nonumber\\& = & \frac{1}{2\pi} \int_0^\theta
\frac{du}{\sqrt{\cos u - \cos \theta}} \int_0^\varepsilon
K(u,v)\,dv \int_v^\varepsilon
\frac{h(\alpha)\,\sin\alpha\,d\alpha}{\sqrt{\cos v - \cos
\alpha}}, \label{eq:66}
\end{eqnarray}
where
 \beq
\label{eq:67} K(u,v)& =& \frac{2}{\pi}\sum_{n=0}^\infty H_n
\cos\left(n+\frac{1}{2}\right)u\cos\left(n+\frac{1}{2}\right)v\\
&&\nonumber\\
& = & -\frac{\cos\frac{1}{2}(v+u)}{2\pi}
\log2\left|\sin\frac{1}{2}(v+u)\right|\nonumber\\
&&\nonumber\\
&& -\frac{\cos\frac{1}{2}(v-u)}{2\pi}
\log2\left|\sin\frac{1}{2}(v-u)\right| \nonumber \\
&&\nonumber\\
& & + \frac{v+u-\pi}{4\pi}\sin\frac{1}{2}(v+u) +
\frac{v-u-\pi}{4\pi}\sin\frac{1}{2}(v-u).\nonumber
 \eeq
Similarly,
\begin{eqnarray}
&& \frac{1}{2}\int_\varepsilon^\pi G(\alpha) \sum_{n=0}^\infty H_n
P_n(\cos \alpha) P_n(\cos \theta) \sin \alpha \, d\alpha
\nonumber  \\
&&\nonumber\\ & = & \frac{1}{2\pi}\int_\varepsilon^\pi
G(\alpha)\sin \alpha \, d\alpha \int_0^\alpha \frac{dv}{\sqrt{\cos
v - \cos \alpha}}
\int_0^\theta \frac{K(u,v)\,du}{\sqrt{\cos u - \cos \theta}}  \nonumber \\
&&\nonumber\\ & = & \frac{1}{2\pi} \int_0^\theta
\frac{du}{\sqrt{\cos u - \cos \theta}} \int_\varepsilon^\pi
G(\alpha)\sin \alpha \, d\alpha \int_0^\alpha
\frac{K(u,v)\,dv}{\sqrt{\cos v - \cos \alpha}}. \label{eq:68}
\end{eqnarray}
Substituting equations (\ref{eq:52}), (\ref{eq:53}),
(\ref{eq:66}), and (\ref{eq:68}) into equation (\ref{18}) yields
\begin{eqnarray}
0 & = & \frac{1}{2\pi}\int_0^\theta \frac{du}{\sqrt{\cos u - \cos
\theta}} \int_u^\varepsilon \frac{h(\alpha) \sin
\alpha\,d\alpha}{\sqrt{\cos u - \cos
\alpha}} \nonumber \\
&&\nonumber\\
& & + \frac{1}{2\pi} \int_0^\theta \frac{du}{\sqrt{\cos u - \cos
\theta}} \int_0^\varepsilon K(u,v)\,dv \int_v^\varepsilon
\frac{h(\alpha)\,\sin\alpha\,d\alpha}{\sqrt{\cos v - \cos \alpha}}
\nonumber \\
&&\nonumber\\
& & + \frac{1}{2\pi} \int_0^\theta \frac{du}{\sqrt{\cos u - \cos
\theta}} \int_\varepsilon^\pi \frac{G(\alpha) \sin \alpha \,
d\alpha}{\sqrt{\cos u - \cos \alpha}} \nonumber \\
&&\nonumber\\
& & + \frac{1}{2\pi} \int_0^\theta \frac{du}{\sqrt{\cos u - \cos
\theta}} \int_\varepsilon^\pi G(\alpha)\sin \alpha \, d\alpha
\int_0^\alpha  \frac{K(u,v)\,dv}{\sqrt{\cos v - \cos
\alpha}},\nonumber
\end{eqnarray}
which is again an Abel-type integral equation. Inverting the Abel
transform \cite{Whittaker}, we obtain
\begin{eqnarray}
 0 & = & \frac{1}{2\pi} \int_u^\varepsilon
\frac{h(\alpha) \sin \alpha\,d\alpha}{\sqrt{\cos u - \cos \alpha}}
+ \frac{1}{2\pi} \int_0^\varepsilon K(u,v)\,dv \int_v^\varepsilon
\frac{h(\alpha)\,\sin\alpha\,d\alpha}{\sqrt{\cos v - \cos \alpha}}
\nonumber \\
&&\label{eq:h-g-K}\\
& & + \frac{1}{2\pi} \int_\varepsilon^\pi \frac{G(\alpha) \sin
\alpha \, d\alpha}{\sqrt{\cos u - \cos \alpha}} + \frac{1}{2\pi}
\int_\varepsilon^\pi G(\alpha)\sin \alpha \, d\alpha \int_0^\alpha
\frac{K(u,v)\,dv}{\sqrt{\cos v - \cos \alpha}}.\nonumber
\end{eqnarray}
Setting
\begin{equation}
H(u) = \frac{1}{2\pi}\int_u^\varepsilon \frac{h(\alpha) \sin
\alpha\,d\alpha}{\sqrt{\cos u - \cos \alpha}},\label{AbelH}
\end{equation}
we invert the Abel transform (\ref{AbelH}) to obtain
\begin{equation}
\label{eq:h} h(\theta) = -\frac{2}{\sin \theta} \frac{d}{d\theta}
\int_{\theta}^\varepsilon \frac{\sin u H(u)\,du}{\sqrt{\cos \theta
- \cos u}}.
\end{equation}
Writing
\begin{equation}
\label{eq:J} J(u) = H(u) + G(u,\varepsilon),
\end{equation}
equation (\ref{eq:h-g-K}) becomes
\begin{equation}
\label{eq:Fredholm} J(u) + \int_0^\varepsilon K(u,v)J(v)\,dv =
M(u),
\end{equation}
where the free term $M(u)$ is given by
\begin{equation}
\label{eq:M} M(u) = -\int_{\varepsilon}^{\pi} K(u,v)G(v,v)\,dv.
\end{equation}
Equation (\ref{eq:Fredholm}) is a Fredholm integral equation for
$J$.
\subsubsection{The second term and the remaining error: $L^2$
estimates}
\label{sec:l2} Equations (\ref{eq:b_0}), (\ref{eq:int-G}), and
(\ref{eq:J}) give that
\begin{equation}
\label{eq:b_0-J} b_0+\frac{2R^2}{3} =
\frac{2R^2}{3}\frac{\pi}{\varepsilon + \sin \varepsilon} +
\frac{\sqrt{2}\pi}{\varepsilon+\sin \varepsilon}\int_0^\varepsilon
J(u) \cos \frac{u}{2}\, du,
\end{equation}
where $J$ is the solution of the Fredholm equation
(\ref{eq:Fredholm}). In this section we show that
$$\frac{\sqrt{2}\pi}{\varepsilon+\sin \varepsilon}\int_0^\varepsilon
J(u) \cos \frac{u}{2}\, du= \left(b_0+\frac{2R^2}{3}
\right)\left(\varepsilon \log \frac{1}{\varepsilon} +
O(\varepsilon) \right),$$ therefore the last term in
eq.(\ref{eq:b_0-J}) should be considered a small correction to the
leading order term $\displaystyle
\frac{R^2}{3}\frac{\pi}{\varepsilon}$, obtained in Section
\ref{Hn0}. This confirms the intuitive results of \cite{Berez},
\cite{Weiss} and gives an estimate on the error term.
Due to the logarithmic singularity of the function $K(u,v)$ (see
(\ref{eq:67})) the operator $K$, defined by
\begin{equation}
Kf (u) = \int_0^\varepsilon K(u,v)f(v)\,dv,
\end{equation}
maps  $L^2[0,\varepsilon]$ into $L^2[0,\varepsilon]$. In Appendix
\ref{ap:K} we derive the estimate
\begin{equation}
\label{eq:L_2-K1} \|K\|_2 \leq \frac{\sqrt{30}}{2\pi}\,\varepsilon
\log \frac{1}{\varepsilon},
\end{equation}
for $\varepsilon \ll 1$. Better estimates can be found; however we
settle for this rough estimate that suffices for our present
purpose.
\subsubsection{Estimate of $\|J\|_2$} In terms of the operator
$K$, equation (\ref{eq:Fredholm}) can be written as
\begin{equation}
J = M - KJ.
\end{equation}
The triangle inequality yields
\begin{equation}
\|J\|_2 \leq \|M\|_2 + \|KJ\|_2 \leq \|M\|_2 + \|K\|_2 \|J\|_2,
\end{equation}
which together with the estimate (\ref{eq:L_2-K1}) gives
\begin{equation}
\label{eq:J-bound} \|J\|_2 \leq \frac{\|M\|_2}{1-\|K\|_2} \leq
\left(1+\varepsilon \log \frac{1}{\varepsilon}\right)
\|M\|_2\quad\mbox{for}\quad\varepsilon \ll 1.
\end{equation}
\subsubsection{Estimate of $\|M\|_2$}
We proceed to find an estimation for $\|M\|_2$. First, we prove
that the kernel satisfies the identity
\begin{equation}
\label{eq:K-0} \int_0^\pi K(u,v)\cos \frac{v}{2}\,dv = 0, \quad
\mbox{for all } u.
\end{equation}
Indeed, by changing the order of summation and integration, we
obtain
\begin{eqnarray}
\int_0^\pi K(u,v)\cos \frac{v}{2}\,dv & = &
\frac{1}{\pi} \sum_{n=1}^\infty \frac{1}{n}\cos
\left(n+\frac{1}{2} \right)u \int_0^\pi \cos\left(n+\frac{1}{2}
\right)v\cos \frac{v}{2}\,dv
\nonumber \\
&&\nonumber \\
& = & \frac{1}{2\pi} \sum_{n=1}^\infty \frac{1}{n}\cos
\left(n+\frac{1}{2} \right)u \int_0^\pi \left(\cos(n+1)v+\cos nv
\right)\,dv
\nonumber \\
&&\nonumber \\
& = & 0.
\end{eqnarray}
Equations (\ref{eq:G}), (\ref{eq:M}), and (\ref{eq:K-0}) imply
that
\begin{equation}
\label{eq:M1} M(u) =
\frac{\sqrt{2}}{\pi}\left(\frac{2R^2}{3}+b_0\right)\int_0^\varepsilon
K(u,v) \cos\frac{v}{2}\,dv.
\end{equation}
The estimate (\ref{eq:L_2-K1}) gives
\begin{equation}
\label{eq:M-bound} \|M\|_2 \leq
\frac{\sqrt{2}}{\pi}\left(\frac{2R^2}{3}+b_0\right) \|K\|_2
\sqrt{\varepsilon} \leq
\frac{\sqrt{15}}{\pi^2}\left(\frac{2R^2}{3}+b_0\right)
\varepsilon^{3/2}\log\frac{1}{\varepsilon}.
\end{equation}
Combining the estimates (\ref{eq:J-bound}) and (\ref{eq:M-bound}),
we obtain for $\varepsilon \ll 1$
\begin{equation}
\|J \|_2 \leq \frac{4}{\pi^2} \left(\frac{2R^2}{3}+b_0 \right)\,
\varepsilon^{3/2}\log\frac{1}{\varepsilon}=
\left(\frac{2R^2}{3}+b_0 \right)\,
O(\varepsilon^{3/2}\log\varepsilon ).\label{eq:J-bound2}
\end{equation}
\subsubsection{The second term and error estimate} The Cauchy-Schwartz inequality implies
that
\begin{equation}
\frac{\sqrt{2}\pi}{\varepsilon+\sin \varepsilon}
\left|\int_0^\varepsilon J(u) \cos \frac{u}{2}\,du \right| \leq
\left(\frac{2R^2}{3}+b_0 \right)\varepsilon \log
\frac{1}{\varepsilon},
\end{equation}
for $\varepsilon \ll 1$, which together with (\ref{eq:b_0-J})
gives
\begin{equation}
\label{eq:b-final} b_0 = \frac{\pi
R^2}{3\varepsilon}\left(1+O(\varepsilon \log\varepsilon )\right) =
\frac{|\Omega|}{4a}\left(1+O(\varepsilon \log\varepsilon )\right).
\end{equation}
To obtain the explicit expression for the term $O(\varepsilon \log
\varepsilon)$, we write the Fredholm integral equation
(\ref{eq:Fredholm}) as
\begin{equation}
(I+K)J = M.
\end{equation}
The estimate (\ref{eq:L_2-K1}) implies that $\|K\|_2 < 1$ for
sufficiently small $\varepsilon$, hence
\begin{equation}
J = M + O\left(\|K\|_2\|M\|_2 \right).
\end{equation}
Thus, using equation (\ref{eq:M1}) and the estimates
(\ref{eq:L_2-K1}) and (\ref{eq:M-bound}), we write the last term
in equation (\ref{eq:b_0-J}) as
\begin{eqnarray}
&&\int_0^\varepsilon J(u)\cos\frac{u}{2}\,du=\int_0^\varepsilon
M(u)\cos \frac{u}{2}\,du + O\left(\varepsilon\|K\|_2\|M\|_2
\right)=\\
&&\nonumber\\
 && \frac{\sqrt{2}}{\pi}\left(b_0 + \frac{2R^2}{3}
\right)\left[\int_0^\varepsilon \int_0^\varepsilon K(u,v)\cos
\frac{u}{2} \cos \frac{v}{2}\,du\,dv + O\left(\varepsilon^3 \log^2
\varepsilon \right)\right].\nonumber
\end{eqnarray}
Equation (\ref{eq:67}) gives the double integral as
 \beqq
\int_0^\varepsilon \int_0^\varepsilon K(u,v)\cos \frac{u}{2} \cos
\frac{v}{2}\,du\,dv = \frac{1}{\pi} \varepsilon^2 \log
\frac{1}{\varepsilon} + O(\varepsilon^2),
 \eeqq
hence
 \beqq
\frac{\sqrt{2}\pi}{\varepsilon+\sin \varepsilon}
\int_0^\varepsilon J(u) \cos \frac{u}{2}\,du = \left(b_0 +
\frac{2R^2}{3} \right) \left[\varepsilon \log
\frac{1}{\varepsilon} + O(\varepsilon)\right].
 \eeqq
Now it follows from equation (\ref{eq:b_0-J}) that
 \beq
b_0 = \frac{|\Omega|}{4a} \left[1+ \varepsilon \log
\frac{1}{\varepsilon} + O(\varepsilon) \right].\label{b0}
 \eeq
\subsection{The MFPT}
Using the explicit expression (\ref{b0}), we obtain the MFPT from
the center of the ball as
\begin{equation}
v\bigg|_{r=0} = u\bigg|_{r=0} + \frac{R^2}{6} = b_0 +
\frac{R^2}{6} = \frac{|\Omega|}{4a}\left[1+\varepsilon \log
\frac{1}{\varepsilon} + O(\varepsilon) \right].
\end{equation}
This is also the averaged MFPT for a uniform initial distribution,
 \beqq
E\tau = \frac{1}{|\Omega|} \int_0^{2\pi}\,d\phi \int_0^\pi \sin
\theta \,d\theta \int_0^R v(r,\theta) r^2 \,dr =
\frac{|\Omega|}{4a}\left[1+\varepsilon \log \frac{1}{\varepsilon}
+ O(\varepsilon) \right].
 \eeqq
\section{Summary and applications}\label{Summary}
The narrow escape problem for a Brownian particle leads to a
singular perturbation problem for a mixed Dirichlet-Neumann
(corner) problem with large Neumann part and small Dirichlet part
of the boundary. The corner problem, that arises in classical
electrostatics (e.g., the electrified disk), elasticity (punch
problems), diffusion and conductance theory, hydrodynamics,
acoustics, and more recently in molecular biophysics, was solved
hitherto mainly for special geometries. In this paper, we have
constructed a leading order asymptotic approximation to the MFPT
in the narrow escape problem for a general smooth domain and have
derived a second term and an error estimate for the case of a
sphere. Our derivation makes Lord Rayleigh's qualitative
observation into a quantitative one. Our leading order analysis of
the general case uses the singularity property of the Neumann
function for a general domain in $\rR^3$. The special case of the
sphere is analyzed by a method developed by Collins and yields a
better result. A different approach to the calculation of the MFPT
would be to use singular perturbation techniques. The vanishing
escape time at the boundary would then be matched to the large
outer escape time of order $\varepsilon^{-1}$ by constructing a
boundary layer near the boundary. The analysis of the MFPT to a
small window at an isolated singular point of the boundary is
postponed to a future paper. Brownian motion through narrow
regions controls flow in many non-equilibrium systems, from
fluidic valves to transistors and ion channels, the protein valves
of biological membranes \cite{Hille}. Indeed, one can view an ion
channel as the ultimate nanovalve---nearly picovalve---in which
macroscopic flows are controlled with atomic resolution. In this
context, the narrow escape problem appeared in the calculation of
the equilibration time of diffusion between two chambers connected
by a capillary \cite{Weiss}. The equilibration time is the
reciprocal of the first eigenvalue of the Neumann problem in this
domain, which depends on the MFPT of a Brownian motion in each
chamber to the narrow connecting channel. The first eigenfunction
is constructed by piecing together the eigenfunctions of the
narrow escape problem in each chamber and in the channel so that
the function and the flux are continuous across the connecting
interfaces. It was assumed in \cite{Weiss} that the flux profile
in the connecting hole was uniform. The structure of the flux
profile, which is proportional to $(a^2-\rho^2)^{-1/2}$, has been
observed by Rayleigh in 1877 \cite{Rayleigh}. Rayleigh first
assumed a radially uniform profile of flux and then refined the
profile of flux going through the channel, allowing it to vary
with the radial distance from the center of the cross section of
the channel, so as to minimize the kinetic energy. A calculation
of the equilibration time was carried out in \cite{Kelman} by
solving the same problem, and gave a result that differs from that
of \cite{Rayleigh}, which was obtained by heuristic means, by less
than two percent. A different approximation, based on the
Fourier-Bessel representation in the pore, was derived in
\cite{Fabrikant2}. Another application of the narrow escape
problem concerns ionic channels \cite{Hille}, and particularly
particle simulations of the permeation process
\cite{Im1}-\cite{Trudy1} that capture much more detail than
continuum models. Up to now, computer simulations are inefficient
because an ion takes so long even to enter a channel and then so
many of the ions return from where they came. From the present
analysis, it becomes clear why ions take so long to enter the
channel. According to (\ref{MR}) the mean time between arrival of
ions at the channel is
 \beq\label{fbr}
 \bar\tau=\frac{E\tau}{N}=\frac1{4DaC},
 \eeq
where $N$ is the number of ions in the simulation and $C$ is their
concentration. A coarse estimate of $\bar\tau$ at the biological
concentration of 0.1Molar, channel radius $a=20\AA$, diffusion
coefficient $D=1.5\times10^{-9}m^2/sec$ is $\bar\tau\approx1nsec$.
In a Brownian dynamics simulation of ions in solution with time
step which is 10 times the relaxation time of the Langevin
equation to the Smoluchowski (diffusion) equation at least 1000
simulation steps are needed on the average for the first ion to
arrive at the channel. It should be taken into account that most
of the ions that arrive at the channel do not cross it \cite{EKS}.

The narrow escape problem comes up in problems of the escape from
a domain composed of a big subdomain with a small hole, connected
to a thin cylinder (or cylinders) of length $L$. If ions that
enter the cylinder do not return to the big subdomain, the MFPT to
the far end of the cylinder is the sum of the MFPT to the small
hole and the MFPT to the far end of the narrow cylinder. The
latter can be approximated by a one-dimensional problem with one
reflecting and one absorbing endpoint. If the domain has a volume
$V$, the approximate expression for the MFPT is
 \beq \label{form}
 E\tau\approx\frac{V}{4\varepsilon D} +\frac{L^2}{2D}.
 \eeq
This method can be extended to a domain composed of many big
subdomains with small holes connected by narrow cylinders. The
case of one sphere of volume $V=\ds{\frac{4\pi R^3}{3}}$, with a
small opening of size $\varepsilon$ connected to a thin cylinder
of length $L$ is relevant in biological micro-structures, such as
dendritic spines in neurobiology. Indeed, the mean time for
calcium ion to diffuse from the spine head to the parent dendrite
through the neck controls the spine-dendrite coupling \cite{PNAS}.
This coupling is involved in the induction of processes such as
synaptic plasticity \cite{Malenka}. Formula (\ref{form}) is useful
for the interpretation of experiments and for the confirmation of
the diffusive motion of ions from the spine head to the dendrite.

Another significant application of the narrow escape formula is to
provide a new definition of the forward binding rate constant in
micro-domains \cite{HS}. Indeed, the forward chemical constant is
really  the flux of particles to a given portion of the boundary,
depending on the substrate location. Up to now, the forward
binding rate was computed using the Smoluchowski formula, which
corresponds to the absorption flux of particles in a given sphere
immersed in an infinite medium. The formula applies when many
particles are involved. But to model chemical reactions in
micro-structures, where a bounded domain contains only a few
particles that bind to a given number of binding sites, the
forward binding rate,
 \[
 k_{\mbox{forward}}=\frac1{\bar\tau},
  \]
has to be computed with $\bar\tau$ given in eq.(\ref{fbr}).

\appendix
\section{Estimate of $\|K\|_2$}\label{ap:K}
\subsection{Estimate of the kernel}
A rough estimate of the kernel, for $0\leq u,v\leq\varepsilon$, is
obtained from equation (\ref{eq:67}) as
 \beqq
K^2(u,v) &\leq&  \frac{5}{4\pi^2} \cos \frac{1}{2}(v+u)\left(
\log 2\left|\sin\frac{1}{2}(v+u)\right|\right)^2\nonumber\\
&&\nonumber\\
&+& \frac{5}{4\pi^2} \cos \frac{1}{2}(v-u) \left(\log
2\left|\sin\frac{1}{2}(v-u)\right|\right)^2.
 \eeqq
Furthermore,
\begin{eqnarray*}
&& \int_0^\varepsilon \cos \frac{1}{2}(v+u) \left( \log
2\left|\sin\frac{1}{2}(v+u)\right|\right)^2 \,du =
\int_{2\sin\frac{1}{2}v}^{2\sin
\frac{1}{2}(v+\varepsilon)} \left(\log x\right)^2\,dx \nonumber \\
&&\\ &  &\leq
2\left(\sin\frac{1}{2}(v+\varepsilon)-\sin\frac{1}{2}v
\right)\left( \log 2\left|\sin\frac{1}{2}v\right|\right)^2 \leq
\varepsilon \cos\frac{1}{2}v
\left(\log2\left|\sin\frac{1}{2}v\right|\right)^2
\end{eqnarray*}
and
\begin{eqnarray*}
\int_0^\varepsilon \varepsilon \cos\frac{1}{2}v \left(\log
2\sin\frac{1}{2}v\right)^2 \,dv & = & \varepsilon
\int_0^{2\sin\frac12\varepsilon} \left(\log x\right)^2 \,dx  \leq
2\varepsilon^2\log^2 \varepsilon. \nonumber
\end{eqnarray*}
Similarly,
\begin{eqnarray*}
 &&\int_0^\varepsilon \cos \frac{1}{2}(v-u) \left(\log\left|2\sin \frac{1}{2}
(v-u)\right|\right)^2 \,dv =\\
&&\\
&& \int_0^{2\sin \frac{1}{2}u}\left(\log x\right)^2 \,dx +
\int_0^{2\sin\frac{1}{2}(\varepsilon-u)}\left(\log x\right)^2\,dx \leq \\
&&\nonumber\\
 &&  2u\log^2 u +
2(\varepsilon-u)\log^2(\varepsilon-u).
\end{eqnarray*}
It follows that
 \beqq \int_0^\varepsilon \left(2u\log^2 u +
2(\varepsilon-u)\log^2(\varepsilon-u)\right)\,du \leq
4\varepsilon^2 \log^2\varepsilon,
 \eeqq
because $u\log u$ is an increasing function in the interval $0
\leq u \leq e^{-2}$. Altogether, we obtain
\begin{eqnarray}
\label{eq:L_2-K} \|K\|_2 & \leq &
\frac{\sqrt{30}}{2\pi}\,\varepsilon \log
\frac{1}{\varepsilon}\quad\mbox{for}\quad \varepsilon \ll e^{-2},
\end{eqnarray}
which is (\ref{eq:L_2-K1}).
\section{Elliptic hole} \label{ap:lure}
We present here, for completeness, Lure's \cite{Lure} solution to
the integral equation (\ref{eq:integral-equation}) in the elliptic
hole case. We define for $\y =(x,y)$
 \beqq
 L(\y)=1-\frac{x^2}{a^2}-\frac{y^2}{b^2}\quad(b\leq a)
 \eeqq
and introduce polar coordinates in the ellipse $\p\Omega_a$
 \beqq
\x=\y+(\rho\cos\theta,\rho\sin\theta),
 \eeqq
with origin at the point $\y$. The integral in
eq.(\ref{eq:integral-equation}) takes the form
\begin{equation}
\int_{\p\Omega_a} \frac{g_0(\x )}{|\x -\y |}\,dS_x =
\int_0^{2\pi}\,d\theta \int_0^{\rho_0(\theta)} \frac{\tilde
g_0\,d\rho}{\sqrt{L(\x )}},
\end{equation}
where $\rho_0(\theta)$ denotes the distance between $\y $ and the
boundary of the ellipse in the direction $\theta$. Expanding
$L(\x)$ in powers of $\rho$, we find that
\begin{equation}
\label{eq:rho-expansion} L(\x
)=1-\frac{(x+\rho\cos\theta)^2}{a^2}-\frac{(y+\rho\sin\theta)^2}{b^2}=
L(\y )-2\phi_1\rho - \phi_2 \rho^2,
\end{equation}
where $ \phi_1
=\ds{\frac{x\cos\theta}{a^2}}+\ds{\frac{y\sin\theta}{b^2}}$ and
$\phi_2 =\ds{
\frac{\cos^2\theta}{a^2}}+\ds{\frac{\sin^2\theta}{b^2}}$. Solving
the quadratic equation (\ref{eq:rho-expansion}) for $\rho$, taking
the positive root, we obtain
\begin{equation}
\rho(\x ) = \frac{1}{\phi_2}\left\{-\phi_1 +
\left[\phi_1^2+\phi_2\left(L(\y )-L(\x )\right) \right]^{1/2}
\right\},
\end{equation}
therefore, for fixed $\y $ and $\theta$,
\begin{equation}
d\rho(\x ) = -\frac{1}{2}\frac{dL(\x
)}{\left[\phi_1^2+\phi_2\left(L(\y )-L(\x )\right)\right]^{1/2}},
\end{equation}
and the integral takes the form
\begin{eqnarray*}
\int_{\p\Omega_a} \frac{g_0(\x )}{|\x -\y |}\,dS_x &=&
\int_0^{2\pi}\,d\theta \int_0^{\ds{L}(\y )}\frac{1}{2}\frac{dL(\x
)}{\left[\phi_1^2+\phi_2\left(L(\y )-L(\x )\right)\right]^{1/2}}
\frac{\tilde g_0}{\sqrt{L(\x )}} \nonumber \\
&&\\
 & = & \int_0^{2\pi}\,d\theta \int_0^{\ds{L}(\y
)}\frac{1}{2}\frac{\tilde g_0\,dz}{\sqrt{\phi_1^2+\phi_2z}
\sqrt{L(\y )-z}}.
\end{eqnarray*}
Substituting $s=\ds{\frac{z}{L(\y )}}$ and setting
$\psi=\ds{\frac{\phi_1^2}{\phi_2 L(\y )}}$, we find that
\begin{eqnarray*}
&&\int_{\p\Omega_a} \frac{g_0(\x )}{|\x -\y |}\,dS_x =
\int_0^{2\pi}\,d\theta \frac{\tilde
g_0}{2\sqrt{\phi_2}}\int_0^{1}\frac{\,ds}{\sqrt{\psi+s}
\sqrt{1-s}}= \nonumber \\
&&\\ &&  \int_0^{2\pi}\,d\theta \frac{\tilde
g_0}{2\sqrt{\phi_2}}2\arctan\sqrt{\frac{\psi+s}{1-s}}\,\bigg|_0^1=\\
&&\\
& & \int_0^{2\pi} \frac{\tilde
g_0}{2\sqrt{\phi_2}}\left(\pi-2\arctan\sqrt{\psi}
\right)\,d\theta=\\
&& \int_0^{2\pi} \ds{\frac{\tilde
g_0\,d\theta}{2\sqrt{\ds{\frac{\cos^2\theta}{a^2}}+\ds{\frac{\sin^2\theta}{b^2}}}}}
\left(\pi-2\arctan\ds{\frac{\ds{\frac{x\cos\theta}{a^2}}+\ds{\frac{y\sin\theta}{b^2}}}
{\sqrt{\ds{\frac{\cos^2\theta}{a^2}}+\ds{\frac{\sin^2\theta}{b^2}}
L(\y )}}} \right). \nonumber
\end{eqnarray*}
The $\arctan$ term changes sign when $\theta$ is replaced by
$\theta+\pi$, therefore its integral vanishes, and we remain with
\begin{eqnarray}
\int_{\p\Omega_a} \frac{g_0(\x )}{|\x -\y |}\,dS_x &=& \frac{\pi
\tilde g_0}{2} \int_0^{2\pi}
\frac{\,d\theta}{\sqrt{\ds{\frac{\cos^2\theta}{a^2}}+\ds{\frac{\sin^2\theta}{b^2}}}}
\nonumber \\
&&\nonumber\\
 &=& 2\pi b \tilde g_0
\int_0^{\ds{\frac{\pi}{2}}}\frac{d\theta}{\sqrt{1-\ds{\frac{a^2-b^2}{b^2}}\sin^2\theta}}
\nonumber \\
&&\nonumber\\
 & = & 2\pi b \tilde g_0 K(e)\label{eq:K},
\end{eqnarray}
where $K(\cdot)$ is the complete elliptic integral of the first
kind, and $e$ is the eccentricity of the ellipse
\begin{equation}
e = \sqrt{1-\frac{b^2}{a^2}}, \quad (a>b).
\end{equation}
We note that the integral (\ref{eq:K}) is independent of $\y $, so
we conclude that (\ref{eq:flux-elliptic}) is the solution of the
integral equation (\ref{eq:integral-equation}).
\section{A pathological example}\label{Pathological}
We have derived an integral equation for the leading order terms
of the flux and the MFPT in the case where the MFPT increases
indefinitely as the relative area of the hole decreases to zero.
However, the MFPT does not necessarily increase to infinity as the
relative area of the hole decreases to zero. This is illustrated
by the following example. Consider a cylinder of length $L$ and
radius $a$. The boundary of the cylinder is reflecting, except for
one of its bases (at $z=0$, say), which is absorbing. The MFPT
problem becomes one dimensional and its solution is
\begin{equation}
v(z) = Lz-\frac{z^2}{2}.
\end{equation}
Here there is neither a boundary layer nor a constant outer
solution; the MFPT grows gradually with $z$. The MFPT, averaged
against a uniform initial distribution in the cylinder, is $E\tau
= \ds{\frac{L^2}{3}}$ and is independent of $a$, that is, the
assumption that the MFPT becomes infinite is violated.\\

\noindent {\bf Acknowledgment:} This research was partially
supported by research grants from the Israel Science Foundation,
US-Israel Binational Science Foundation, and the NIH Grant No.
UPSHS 5 RO1 GM 067241.

\end{document}